\documentclass{llncs}
\pdfoutput=1
\usepackage{graphicx}

\usepackage{times}
\usepackage{mathptm}
\usepackage{cite}
\usepackage{amsmath}

\DeclareSymbolFont{AMSb}{U}{msb}{m}{n}
\DeclareSymbolFontAlphabet{\Bbb}{AMSb}

\makeatletter
\def\hb@xt@{\hbox to }
\makeatother

\let\oldendproof\endproof
\def\endproof{\qed\oldendproof}

\begin{document}
\title{Graph-Theoretic Solutions\\ to Computational Geometry Problems} 

\author{David Eppstein}

\institute{Computer Science Department, University of California, Irvine}

\maketitle   

\begin{abstract}
Many problems in computational geometry are not stated in graph-theoretic terms, but can be solved efficiently by constructing an auxiliary graph and performing a graph-theoretic algorithm on it. Often, the efficiency of the algorithm depends on the special properties of the graph constructed in this way. We survey the art gallery problem, partition into rectangles, minimum-diameter clustering, rectilinear cartogram construction, mesh stripification, angle optimization in tilings, and metric embedding from this perspective.
\end{abstract}

\section{Introduction}

Graph algorithms and computational geometry form separate communities with separate conferences such as the International Workshop on Graph-Theoretic Concepts in Computer Science and the ACM Symposium on Computational Geometry, respectively, but they also meet in broader algorithms conferences, and there has been much interplay between the research topics in the two areas.

Many classical graph algorithm problems have geometric analogues, algorithmic problems on graphs defined by a geometric input. In most cases, problems of this type can be solved directly by constructing the graph and then applying a general-purpose graph algorithm, but can be sped up by examining the graph algorithm's structure more closely and applying appropriate geometric data structures. A notable instance of this phenomenon is the Euclidean minimum spanning tree (the spanning tree of a complete graph in which the vertices represent points and edge lengths are Euclidean distances): by using a Delaunay triangulation in place of a complete graph, the quadratic time of a naive algorithm can be improved to $O(n\log n)$~\cite{ShaHoe-FOCS-75}. Other work along these lines includes algorithms for Euclidean matching~\cite{AgaEfrSha-SJC-00,Vai-SJC-89} and bipartiteness testing~\cite{Epp-TALG-bgig}.

Graph drawing, the visualization of graphs via geometric graph representations~\cite{BatEadTam-98,JunMut-04,NisRah-04}, forms another community represented by the annual International Symposium on Graph Drawing. Most work in the area concerns drawings in which a graph's vertices are represented as geometric points, disks, or polygons, and its edges are represented as line segments or curves. Researchers in this area seek algorithms that optimize mathematical stand-ins for their aesthetic quality and readability such as the number of crossings, the number of bends in non-straight edges, the angles formed by adjacent edges, the area of the drawing, or the amount of symmetry that the drawing displays; the interplay between these different measures provides much scope for research.

Geometric techniques have sometimes also been applied to solve problems that were originally defined in purely graph-theoretic terms. One instance concerns parametric minimum spanning trees, minimum spanning trees of graphs whose edge weights are linear functions of a parameter. Dey proved~\cite{Dey-DCG-98} that $O(mn^{1/3})$ different trees may be formed in this way from a parametric graph with $m$ edges and $n$ vertices, using the \emph{crossing number inequality} that $s$ simple plane curves with $n$ shared endpoints have $\Omega(s^3/n^2)$ crossings~\cite{AjtChvNew-TaPoC-82,Lei-83}, while the best known lower bound, $\Omega(m\alpha(n))$, involves a reduction from another geometric problem, lower envelopes of line segments~\cite{Epp-DCG-98}.

In this paper we survey connections between computational geometry and graph algorithms of yet another type: problems in computational geometry that, although not initially defined graph-theoretically, may be solved by constructing an auxiliary graph from the input, applying a purely graph-theoretic algorithm to this auxiliary graph, and translating the output of this algorithm back into geometric terms. We discuss problems including the art gallery problem, partition into rectangles, minimum-diameter clustering, rectilinear cartogram construction, mesh stripification, angle optimization in tilings, and metric embedding. The ordering of the problems is roughly chronological, and the selection of topics is idiosyncratic and (especially for the later problems) largely drawn from the author's own research rather than being comprehensive.

\section{Art gallery problems}

\begin{figure}[t]
\center\includegraphics[width=3in]{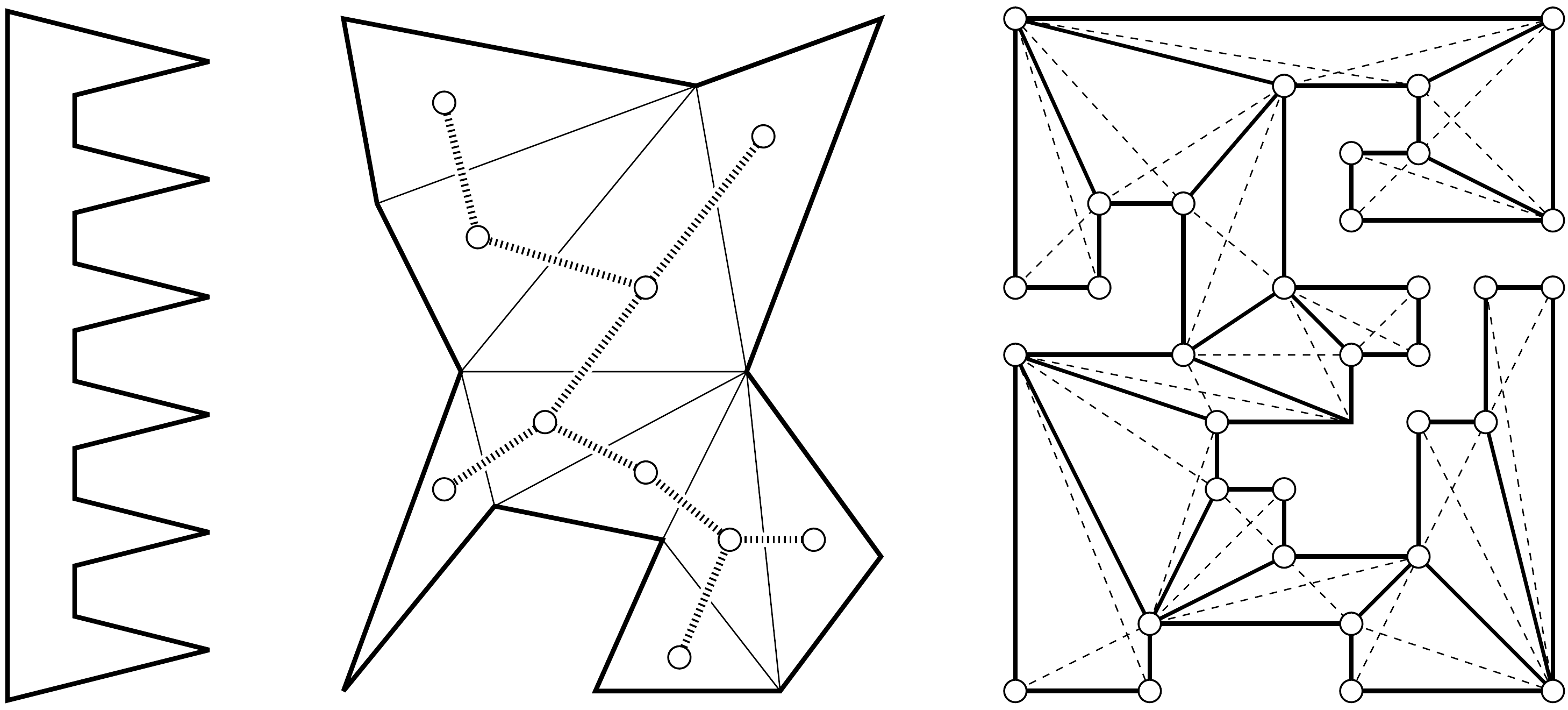}
\caption{The art gallery problem. Left: a comb polygon requiring $\lfloor n/3\rfloor$ guards. Center: triangulating an input polygon produces a maximal outerplanar graph, the weak planar dual of which is a tree. Right: The kinggraph formed by adding diagonals (dashed) to a partition of an orthogonal polygon into convex quadrilaterals.}
\label{fig:ArtGallery}
\end{figure}

Most computer scientists are familiar both with Chv\'atal's \emph{art gallery theorem}~\cite{Chv-JCTB-75,ORo-87} that every $n$-vertex simple polygon has a set of $\lfloor n/3 \rfloor$ \emph{guard points} from which the whole polygon may be seen, and with Fisk's elegant graph-theoretic proof~\cite{Fis-JCTB-78}. One begins the proof by adding a maximal set of non-crossing diagonals to the polygon, partitioning it into triangles. Graph-theoretically, the vertices, sides, and added diagonals of the polygon form a maximal outerplanar graph; the \emph{weak dual} of this graph (the adjacency graph of the triangles, omitting the outer face) is a tree (Figure~\ref{fig:ArtGallery}, center). Every maximal outerplanar graph may be colored with three colors, as may be proven by induction: the result is clearly true when the graph is a triangle, and any larger maximal planar graph may be colored by removing a leaf from the dual tree, coloring the remaining graph, restoring the leaf, and observing that the restored vertex has only two neighbors and therefore has a free color available. Each triangle has one vertex of each color, so each color class forms a valid guard set, and the smallest of the color classes has at most $\lfloor n/3 \rfloor$ vertices. This proof technique translates into an efficient algorithm: polygons may be triangulated in linear time~\cite{Cha-DCG-91}, and the induction proof leads to a linear-time algorithm for 3-coloring maximal planar graphs.

A less well-known variant of the art gallery problem concerns \emph{simple orthogonal polygons}, simple polygons all of whose sides are parallel to the coordinate axes. Simple orthogonal polygons need only $\lfloor n/4\rfloor$ guards, as can be shown again by graph coloring.  Every simple orthogonal polygon can be partitioned by diagonals into convex quadrilaterals~\cite{KahKlaKle-SJDM-83}; the resulting tree of quadrilaterals can be viewed as a special type of \emph{squaregraph}, a planar graph in which every bounded face has four sides and every vertex either belongs to the unbounded face or has at least four incident edges~\cite{CheDraVax-SODA-02,BanCheEpp-09}. Adding the two diagonals of each quadrilateral to a squaregraph forms a \emph{kinggraph}~\cite{CheDraVax-SODA-02} (Figure~\ref{fig:ArtGallery}, right). As with maximal outerplanar graphs, the kinggraphs derived from simple orthogonal polygons may be shown to be 4-chromatic by removing leaves of the dual tree.\footnote{More generally, every kinggraph is 4-chromatic, but a proof is beyond the scope of this survey.} In any four-coloring, each quadrilateral has a vertex of each color, so each color class forms a guard set, and the smallest color class has at most $\lfloor n/4 \rfloor$ guards. This bound is tight: a comb-shaped orthogonal polygon requires  $\lfloor n/4 \rfloor$ guards, just as a comb-shaped simple polygon requiring $\lfloor n/3 \rfloor$ guards shows that the bound for the standard art gallery problem is tight (Figure~\ref{fig:ArtGallery}, left).

The biggest remaining open problem in art gallery theory concerns edge guards: how many edges must one select, from a simple polygon, so that every point of the polygon may be seen from some point on some selected edge? It is not clear, in this case, what graph one should define on the edges of the polygon to force the color classes of a coloring of the graph to form edge guard sets.

\section{Partition into rectangles}

\begin{figure}[t]
\centering\includegraphics[width=4in]{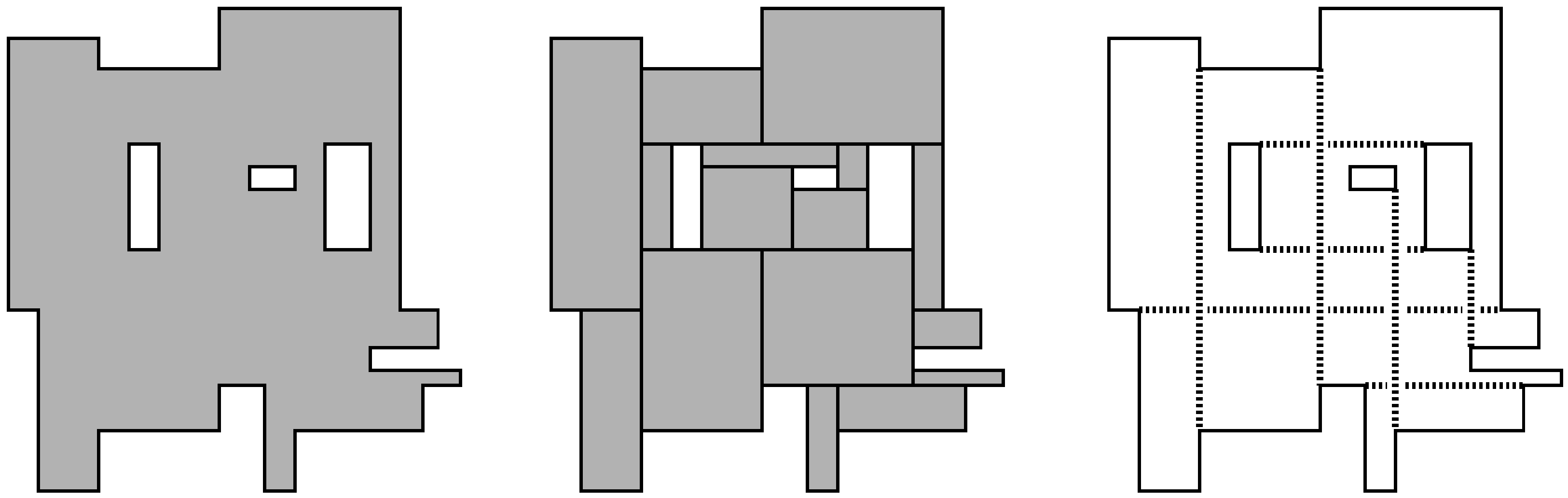}
\caption{Partitioning an orthogonal polygon (left) into the minimum number of rectangles (center). The right figure shows the axis-parallel diagonals that connect pairs of concave vertices; the rectangle partition problem may be solved by finding a maximum independent set in the bipartite intersection graph of these diagonals.}
\label{fig:RectanglePartition}
\end{figure}

Many geometric algorithms take as input a complicated polygonal domain and cover or partition it using simpler shapes~\cite{Kei-HCG-99}; the partitions into triangles and quadrilaterals from the previous section are important examples. Another problem of this type concerns the partition into rectangles of an orthogonal polygon. The input polygon may have polygonal holes, unlike the simple orthogonal polygons of the previous section, and the rectangles are not required to meet edge-to-edge and vertex-to-vertex. The goal is to minimize the total number of rectangles in the partition (Figure~\ref{fig:RectanglePartition}).

Rectangular partitions have many applications. In VLSI design, it is necessary to decompose masks into the simpler shapes available in lithographic pattern generators~\cite{Pat-CAD-77}, and similar mask decomposition problems also arise in DNA microarray design~\cite{HanHubLip-CT-02}. Rectangular partitions can simplify convolution operations in image processing~\cite{FerSanSkl-CVGIP-84} and can be used to compress bitmap images~\cite{CheIyeKas-TSE-88}. Closely related matrix decomposition problems have been applied to radiation therapy planning~\cite{Eng-DAM-09,Kal-EJC-09}, and rectangular partitions have also been used to design robot self-assembly sequences~\cite{LiZha-IROS-05}.

Define a \emph{good diagonal} to be an axis-parallel line segment interior to the input polygon that connects two concave vertices of the polygon. As several authors independently discovered~\cite{FerSanSkl-CVGIP-84,LipLodLuc-FI-79,Oht-ISCAS-82}, the minimum number of rectangles in a partition of a polygon with $n$ vertices and $h$ holes is
$n/2 + h - g - 1$, where $g$ is the maximum size of a set of disjoint good diagonals. To see this, consider the number of corners of rectangles in a partition, four times the number of rectangles. Let $G$ be a maximal set of disjoint good diagonals that form a subset of the line segments in some partition of an orthogonal polygon, and define a \emph{bad vertex} to be a nonconvex polygon vertex that is not an endpoint of $G$. Each polygon vertex forms at least one rectangle corner. Additionally, for each bad vertex $v$, let $s$ be an interior line segment of the partition having $v$ as an endpoint. Then either $s$ crosses a segment in $G$, and the two corners formed by the crossing that are on the same side as $v$ with respect to the crossed segment can be charged to~$v$, or $s$ ends at a non-vertex and the two corners formed at its other endpoint can be charged to~$v$. This charging scheme shows that the number of rectangle corners in the partition is at least $n$ plus twice the number of bad vertices. The number of nonconvex vertices in any orthogonal polygon with $h$ holes is $n/2 + 2h - 2$, so the number of bad vertices is $n/2 + 2h - 2|G| - 2$, and a lower bound of $n/2 + h - |G| - 1$ rectangles follows. Conversely, if $G$ is any set of disjoint good diagonals, a partition with exactly $n/2 + h - |G| - 1$ rectangles may be found by considering the bad vertices for $G$, one at a time, and extending a line segment from each bad vertex to the closest previously-added segment or polygon side. Thus, finding a partition into a minimum number of rectangles is equivalent to finding a maximum number of disjoint good diagonals.

The intersection graph of the good diagonals is bipartite: each horizontal diagonal intersects only vertical diagonals and vice versa. Therefore, finding the maximum number of disjoint good diagonals translates, in graph-theoretic terms, into finding a maximum independent set in a bipartite graph. By K\"onig's theorem~\cite{Kon-MeFL-31}, in any $n$-vertex bipartite graph the maximum independent set has size $n-M$, where $M$ is the cardinality of a maximum matching; an independent set of this size may be found from a maximum matching by partitioning the vertices according to the lengths of the shortest alternating paths from an unmatched vertex to the given vertex, and including the vertices at even levels of this partition. Therefore, the maximum independent set of the intersection graph, the corresponding maximum set of disjoint good diagonals, and a partition into a minimum number of rectangles, may all be found in polynomial time.

A naive implementation of an algorithm that reduces the problem to a bipartite graph and then applies a general-purpose bipartite graph matching algorithm would solve the problem in time $O(n^{2.5})$, where $n$ denotes the number of vertices of the input polygon: this is the time needed to apply the Hopcroft--Karp matching algorithm~\cite{HopKar-SJC-73} to the intersection graph of the good diagonals, which may have $\Theta(n)$ vertices and $\Theta(n^2)$ edges. However, by using geometric range searching data structures to speed up the search for alternating paths within the matching algorithm, it is possible to improve the overall running time to $O(n^{3/2}\log n)$~\cite{Lip-Nw-83,Lip-IPL-84,ImaAsa-SJC-86}.

\section{Minimum diameter clustering}

The \emph{diameter} of a set of points is the maximum distance of any two of its points. The problem of finding low-diameter subsets of larger sets of input points~\cite{AggImaKat-Algs-91,EppEri-DCG-94} may be formulated in several different ways: one may take as input a number $k$ and produce as output the subset of $k$ points with minimum diameter, one may take as input a number $D$ and produce as output the largest subset with diameter at most $D$, or one may solve the \emph{decision problem} in which $D$ and $k$ are both given and the task is to determine whether there exists a set of $k$ points with diameter at most $D$. Since there are only $O(n^2)$ potential diameter values and $n$ different values of $k$, an efficient algorithm for any one of these tasks leads to efficient algorithms for the other two tasks as well.

\begin{figure}[t]
\centering\includegraphics[width=3.5in]{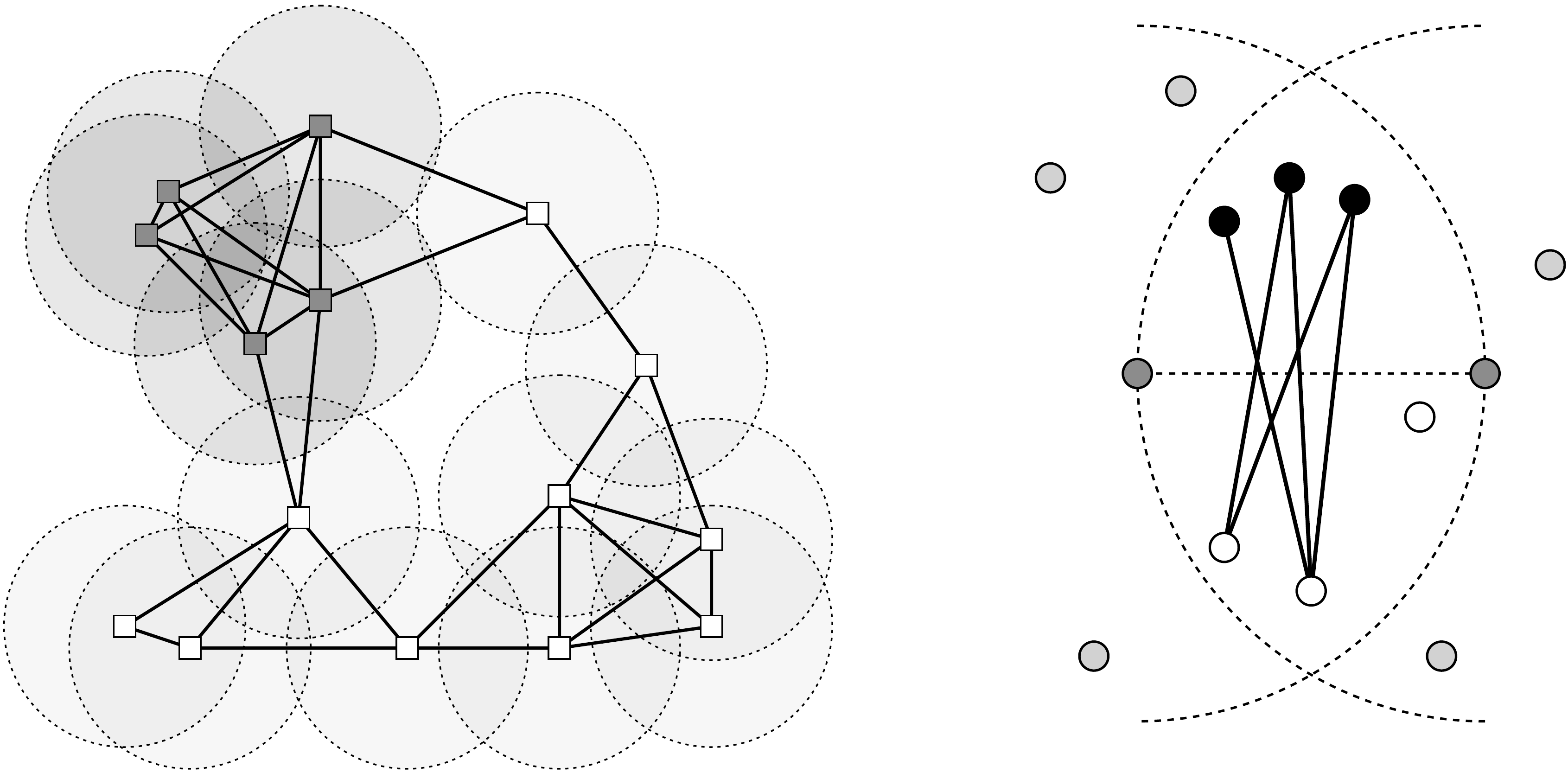}
\caption{Minimum diameter clustering. Left: a unit disk graph and its largest clique (the five darker circles in the upper left). Right: for any two specified points, the largest cluster having those two points as its diametral pair may be found as a maximum independent set of a bipartite graph in which the two sides of the bipartition are two halves of a lune defined by the two specified points.}
\label{fig:MinDiam}
\end{figure}

Finding the largest size $k$ of a set with diameter at most a given value $D$ has a direct translation to a graph theoretic problem, but the very directness of the translation means that it is unhelpful: it provides merely a restatement of the problem rather than conveying any new insight. We may scale the input point set so that $D=2$; then what we seek is the largest clique in a \emph{unit disk graph}, the intersection graph of unit disks centered at the points (Figure~\ref{fig:MinDiam}, left); note that a set of disks forming a clique need not have a common intersection point. Maximum cliques in unit disk graphs may be found in polynomial time, given a disk representation of the graph~\cite{ClaColJoh-DM-90} but this is just a trivial restatement of the minimum diameter clustering problem; it is NP-hard to find a disk representation given only a graph-theoretic description of a unit disk graph~\cite{BreKir-CGTA-98}.

However, just as in the rectangle partition problem, minimum diameter clustering may be reduced to a more basic graph-theoretic problem, maximum independent sets in bipartite graphs~\cite{AggImaKat-Algs-91,ClaColJoh-DM-90}. Suppose that we know or can guess the two points $p$ and $q$ forming the diametral pair in a cluster. Then, all the other points of the cluster must be within the \emph{lune} formed by intersecting the two circles with $pq$ as radius, centered on $p$ and $q$. The largest possible cluster within the lune is the maximum clique of the unit disk graph with disk centers inside this lune, or equivalently the maximum independent set of the complement $G$ of this restricted unit disk graph. But $G$ is bipartite: if the lune is bisected by line $pq$, then a point in $G$ can be connected by an edge only to other points on the other side of the bisection line (Figure~\ref{fig:MinDiam}, right). Therefore, its maximum independent set may be found in polynomial time, as discussed in the previous section.

Based on this idea, one could test all $\Theta(n^2)$ pairs $pq$, find the bipartite graph derived from each pair and apply a bipartite matching algorithm to it, and return the best cluster found from all of these separate tests. However it is more efficient to use dynamic graph algorithms to share work between multiple different matching problems~\cite{EppEri-DCG-94}. Suppose we seek the size of the largest cluster for a given~$D$. For each point $p$ that could be an endpoint of a diametral pair, consider the lune defined by a segment of length $D$ with $p$ as one endpoint. If the defining segment rotates through an angle of $2\pi$ around $p$, the lune rotates with it, and the set of input points inside the lune undergoes a sequence of $O(n)$ discrete changes in which some point joins or leaves the set. After each change, one may update the maximum matching of the bipartite graph defined from the lune by a single alternating path search. Thus, the overall algorithm loops through all $n$ possible choices of $p$, and performs a nested loop through the $O(n)$ set insertion and deletion operations defined by rotating a size-$D$ lune around $p$. For each set update operation the algorithm performs an alternating path search to update a maximum matching and the maximum independent set in the bipartite graph defined from the lune, and when the nested loops terminate the algorithm returns the largest cluster found in each of these searches. As Aggarwal et al.~\cite{AggImaKat-Algs-91} describe, each step of an alternating path search may be performed in logarithmic time with the aid of the circular hull data structure of Hershberger and Suri~\cite{HerSur-Algs-91}. Therefore, each alternating path search takes time $O(n\log n)$, the sequence of $O(n)$ alternating path searches for a single choice of $p$ takes time $O(n^2\log n)$, and the overall clustering algorithm takes time $O(n^3\log n)$~\cite{EppEri-DCG-94}.

If $k$ rather than $D$ is given as input, the problem may be solved by a binary search among the $O(n^2)$ different distances defined by the input points, that checks for each distance whether it is the diameter of some $k$-point cluster. For this variant, the time may be further improved to $O(n\log n+k^2 n\log^2 k)$ by using a $k$-nearest-neighbor calculation to reduce the problem to $O(n/k)$ subproblems with $O(k)$ points per subproblem~\cite{EppEri-DCG-94}.

It appears to be open whether there exist polynomial time algorithms for solving the analogous minimum diameter clustering problems in higher dimensions (equivalently, finding maximum cliques in intersection graphs of unit balls in higher dimensions) or for finding a maximum clique in a unit disk graph when a geometric representation of the graph is not given.

\section{Bend minimization}

A rectilinear cartogram~\cite{Rai-GR-34,KreSpe-CGTA-07,Mum-PhD-08} is a diagram in which geographic regions have been replaced by orthogonal polygons, with approximately the same shapes and in approximately the same positions with respect to each other as they hold geographically, but in which the areas of the regions have been distorted to represent numerical data about the regions unrelated to their physical areas. In introducing these diagrams in 1934, Raisz~\cite{Rai-GR-34} wrote, ``it should be emphasized that the statistical cartogram is not a map,'' and the stylization inherent in using orthogonal polygons helps perform this emphasis.

\begin{figure}[t]
\centering\includegraphics[height=1.5in]{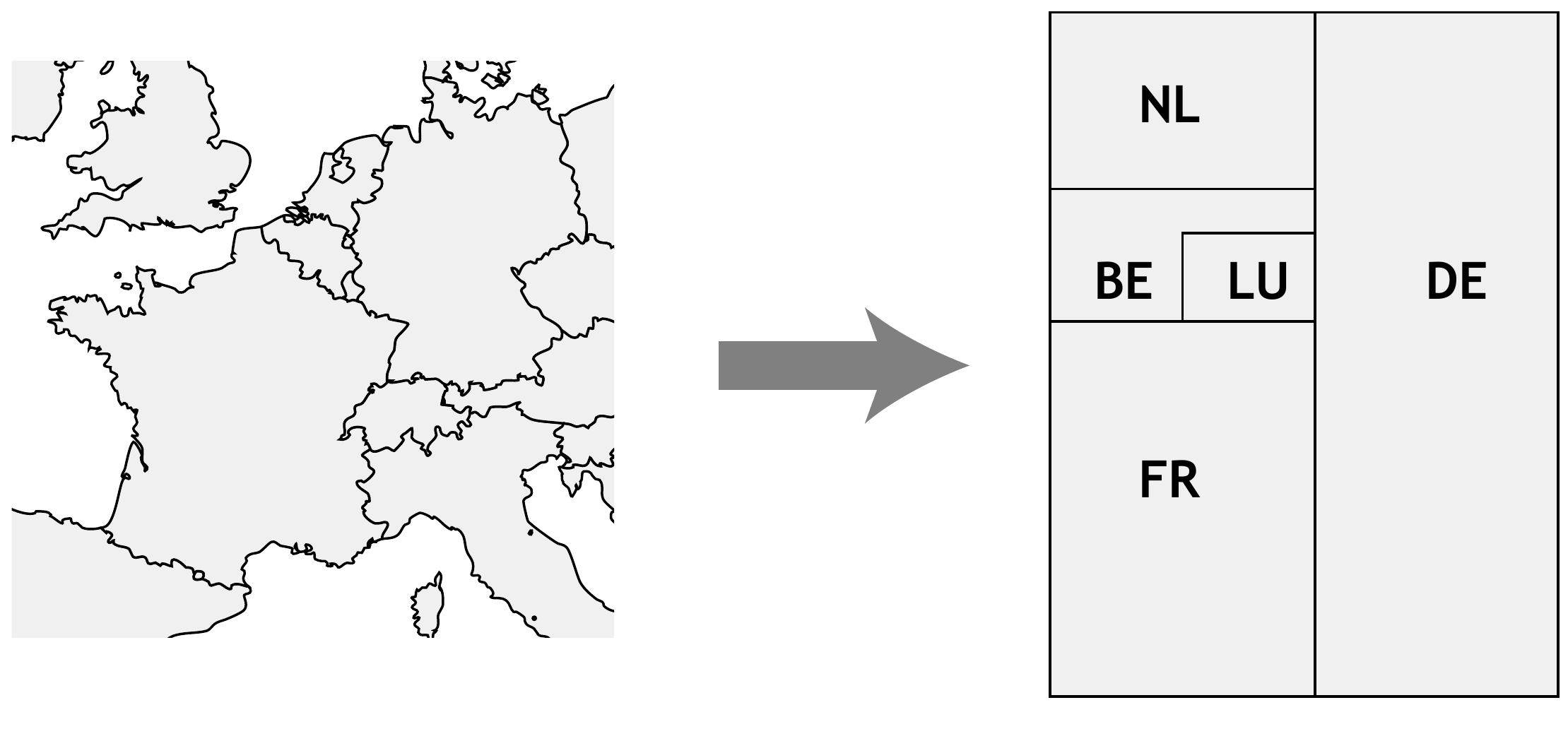}
\caption{A rectilinear cartogram of five countries in northwestern Europe. Map based on a CC-BY-SA licensed image by Brianski, Canuckguy, Zaparojdik, Madman2001, Roke, and Ssolbergj, 
online at http://commons.wikimedia.org/wiki/File:Blank\_Map\_of\_Europe\_-w\_boundaries.svg.}
\label{fig:Cartogram}
\end{figure}

We have studied algorithms for constructing cartograms that can accomodate arbitrary area assignments~\cite{EppMumSpe-SCG-09} and in which the adjacencies between regions have desired orientations~\cite{EppMum-WADS-09} but a more basic problem, constructing a cartogram in which the region shapes are as simple as possible, had already been solved in a different context, that of graph drawing~\cite{Tam-SJC-87}.  Simple shapes such as rectangles aid the viewer in measuring and comparing areas and hence in understanding the data represented by a cartogram. We would like to find a cartogram that represents a given map with a minimum number of \emph{bends} (corners interior to the boundary between two adjacent regions); for instance, in Figure~\ref{fig:Cartogram} there is one bend, between Belgium and Luxembourg; this is optimal, as Luxembourg is surrounded by only three countries and hence must have at least one bend on its boundary. As Tamassia~\cite{Tam-SJC-87} shows, this problem of bend minimization can be solved in polynomial time by translating it into a network flow problem.

To form a flow problem that represents the bends of some given cartogram, construct a graph that has a single ``circulation'' vertex, a vertex for each region of the cartogram (including the exterior of the diagram as one of the regions),  and a vertex for each junction where three or four regions of the cartogram meet (four regions at a junction may be needed to model places like the Four Corners in the southwest of the U.S. where four states meet, but five or more regions cannot meet at a single junction in an orthogonal cartogram). Include edges from the circulation vertex to each other vertex, between each two vertices that represent adjacent regions, and between two vertices that represent an incident junction-region pair. Label this graph with flow amounts, as follows: each junction vertex has four incoming units of flow from the circulation vertex, and sends one unit of flow to the vertices representing the regions in the four quadrants incident to the junction. For each bend in the cartogram, send one unit of flow from the region that has a convex vertex at the bend to the region that has a concave vertex. For each interior region having $k$ junctions on its boundary, send $4-2k$ units of flow to the circulation vertex (or equivalently, if $4-2k<0$, send $2k-4$ units from the circulation vertex to the region vertex). And for an exterior region with $k$ junctions on its boundary, send $2k+4$ units of flow to the circulation vertex. The result can be shown to be a valid circulation: that is, for each vertex of the flow graph, the numbers of incoming and outgoing flow units are equal. For the junction vertices, this is clear because the four incoming units are balanced by the four quadrants into which a unit of flow is sent. An interior region with $k$ junctions and no bends must form a rectangle, as a junction cannot form a concave corner; thus, it has four incident junction vertices that send a single unit of flow, and the remaining $k-4$ incident junctions send two units of flow into the region, for a total of $2k-4$ incoming units, balancing the flow to the circulation vertex. Each additional cave corner at a bend causes one unit of incoming flow from another region, but must be balanced by an additional convex corner; if that convex corner belongs to a bend, it provides a unit of outgoing flow, and if it belongs to a junction then that junction sends one fewer unit of incoming flow, in either case leading to the same total flow balance. A similar argument shows that the flow into and out of the exterior region vertex is also balanced, from which it follows that the flow must be balanced at the single remaining vertex, the circulation vertex.

Conversely, as Tamassia shows, one can assign costs and flow capacities to this network in order to force a minimum cost circulation to correspond to a minimum-bend drawing. Capacity constraints are needed to force the incoming flow to each junction vertex to be exactly four units; the edges between adjacent region vertices are assigned unit cost, and all other edges have zero cost. With these constraints and costs, the flow described above has cost equal to its number of bends. Conversely, any integer solution to the capacitated circulation problem can be translated into a drawing in which the total number of bends is equal to the cost of the circulation, so a cartogram with the minimum number of bends can be found from a minimum-cost circulation, which can be constructed in polynomial time. The flow graph is an \emph{apex graph}: if one vertex, the circulation vertex, is removed, the rest is planar. Therefore, specialized techniques for finding flows in planar graphs may be used to speed up this algorithm~\cite{ImaIwa-SIGAL-90,GarTam-GD-96,Mil-SJC-95}.

\section{Mesh stripification}

\begin{figure}[t]
\centering\includegraphics[height=1.7in]{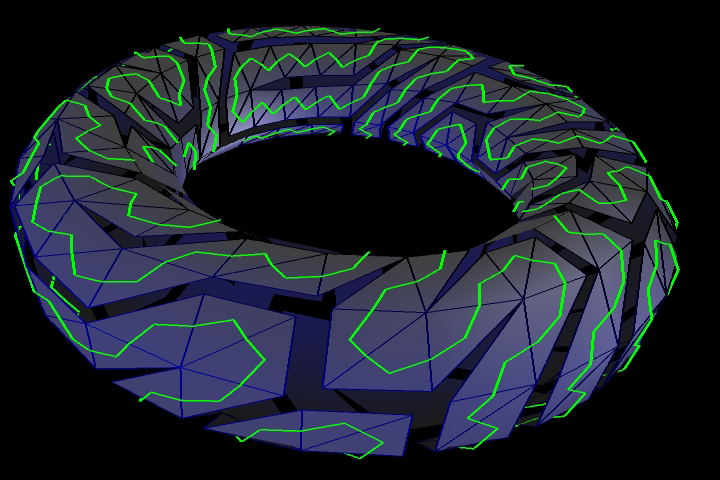}
\qquad\includegraphics[height=1.7in]{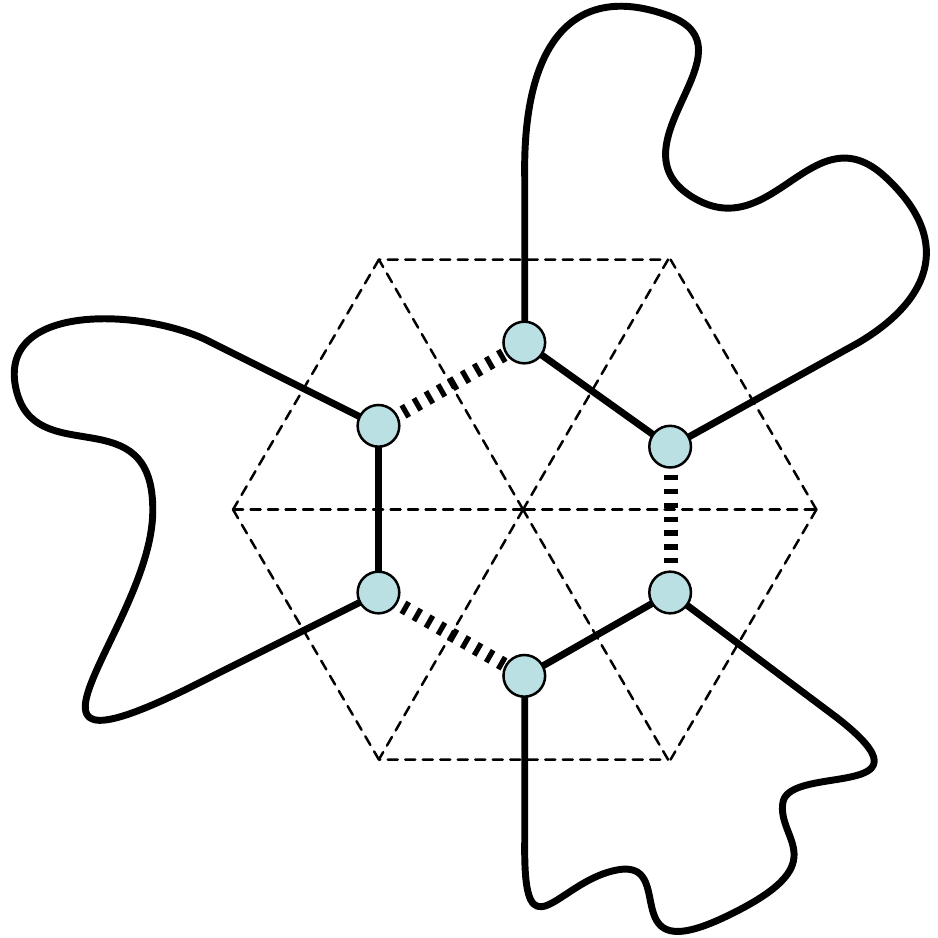}
\caption{Left: a single strip in a triangulated torus, rendered by M. Gopi, from~\cite{Epp-JGAA-07}. Right: a local move at a vertex allows three cycles to be merged into a single cycle.}
\label{fig:Stripification}
\end{figure}

\emph{Stripification} refers to the problem in computer graphics of partitioning a triangulated surface model of a three-dimensional object into \emph{triangle strips}, sequences of triangles meeting edge-to-edge~\cite{ArkHelMit-VC-96,EppGop-EG-04,EvaSkiVar-V-96,XiaHelMit-3DG-99}. Such a partition allows for fast rendering by requiring the coordinates of only one vertex per triangle to be transmitted to the graphics hardware; the other two vertex locations may be found from the previous triangle in the strip. Triangle strips aid in data compression of geometric models: a predicted location that aids in compressing the coordinates of each successive vertex may be found by extrapolation from the previous triangle in a strip~\cite{Dee-SIGGRAPH-95}. Additionally, if a mesh can be represented as a single cyclic triangle strip, as in Figure~\ref{fig:Stripification} (left), the structure of the strip may be used to cover the surface of the model with a space-filling curve~\cite{EppGop-EG-04}, and contractions of the boundary edges of the strip can be used for topology-preserving simplifications of the model~\cite{Dia-PhD-09}.

Finding such a strip may be viewed graph-theoretically, as finding a Hamiltonian cycle in the 3-regular dual graph of the triangulation, a graph with one vertex per triangle and one edge for each pair of adjacent triangles. However, this is problematic for two reasons. First, the Hamiltonian cycle problem is NP-complete and the known exponential-time algorithms for the problem~\cite{Epp-JGAA-07,IwaNak-COCOON-07} are only capable of solving it within a reasonable time for models of at most a few hundred triangles. Second, and more importantly, not all triangulated models, even with spherical topology, have single triangle strips of this type. Tutte's counterexample to Tait's conjecture that 3-regular polyhedra are Hamiltonian~\cite{Tai-PM-84,Tut-JLMS-46} dualizes to become a triangulated mesh that cannot be represented as a single cycle of triangles.

However, it is important to realize that the dual graph does not uniquely represent the geometry: we may change the triangulation, and hence change the graph, without changing the model's shape. The dual graph of any triangulated model is both 3-regular and bridgeless; therefore, by a theorem of Petersen~\cite{Pet-AM-91}, it has a perfect matching, which may be found efficiently~\cite{BieBosDem-Algs-01}. The set of edges complementary to the matching forms a partition of the triangulation into a collection of disjoint cycles~\cite{EppGop-EG-04}; however, there may be more than one of these cycles. As we observed~\cite{EppGop-EG-04}, in many cases a local move at a vertex of the triangulation, that swaps matched and unmatched edges connecting the triangles sharing that vertex, may reduce the number of cycles (Figure~\ref{fig:Stripification}, right). If no such move is available, two adjacent triangles from two different cycles may be bisected across their shared edge, leading to a new triangulation of the same surface with the property that a cycle-reducing local move is available at the newly created vertex. By repeating this subdivision process, one eventually reaches a triangulation that covers a surface identical to the input model, but one that has a different dual graph than the input and that can be represented as a single strip. Although theoretically this could increase the total number of triangles by as much as a factor of $3/2$, in practice we saw at most a 2\% increase. 

\section{Angle optimization of tilings}

\begin{figure}[t]
\includegraphics[height=2.2in]{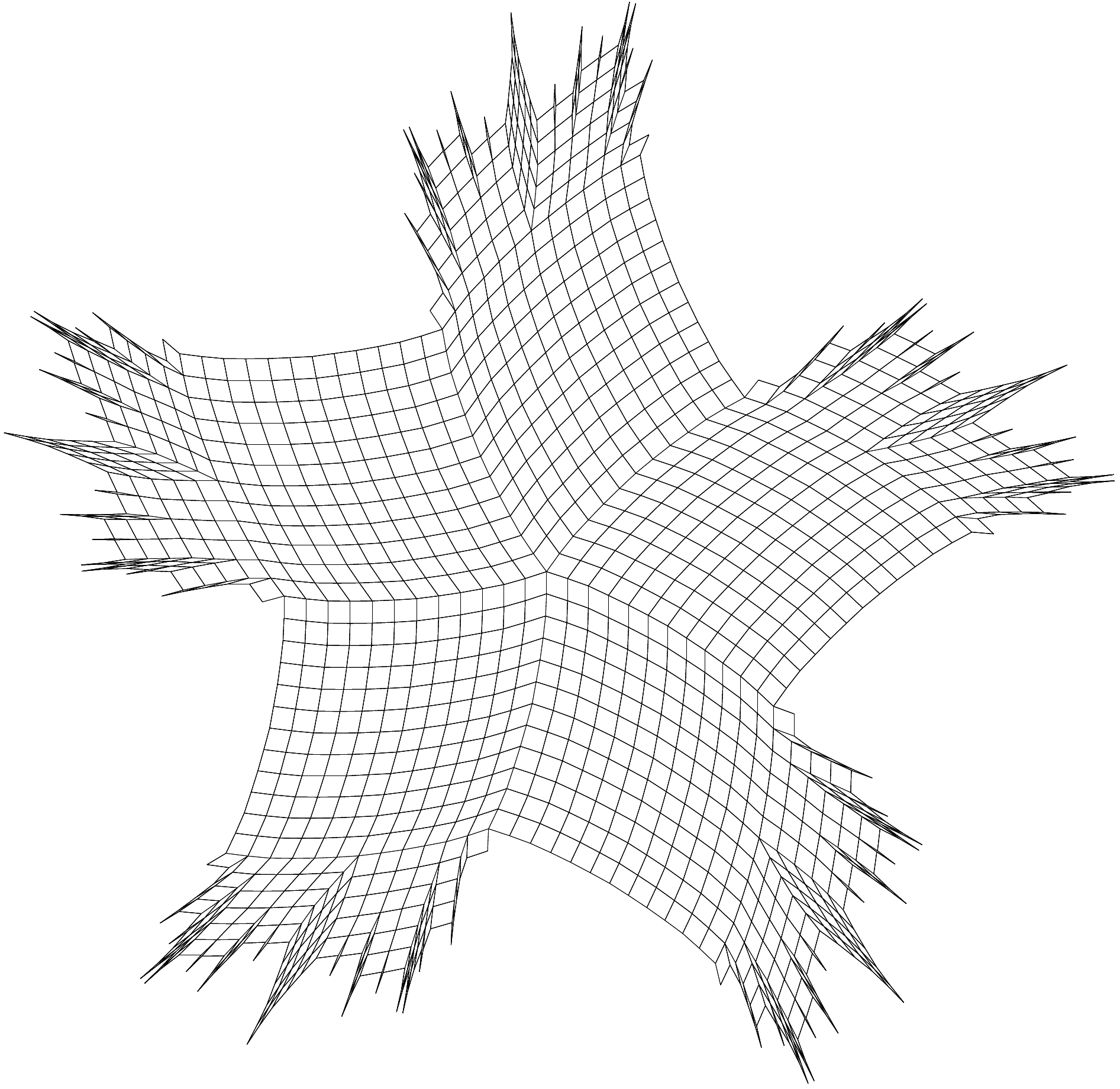}\qquad\includegraphics[height=2.2in]{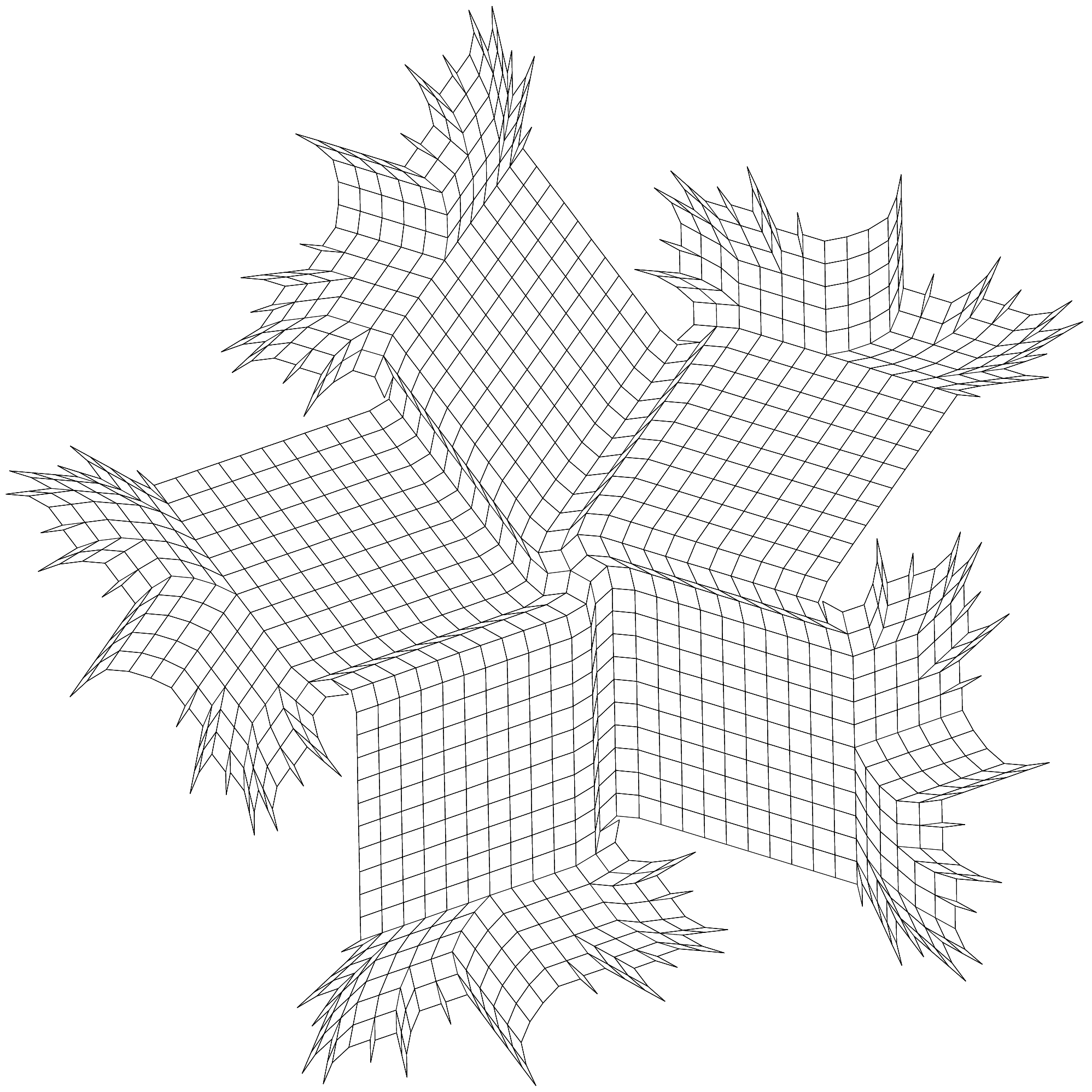}
\caption{Two drawings of the squaregraph dual to a 5-chromatic triangle-free circle graph described by Ageev~\cite{Age-DM-96}, from~\cite{BanCheEpp-09}.}
\label{fig:Ageev}
\end{figure}

The planar dual of any arrangement of lines in the plane may be represented as a \emph{tiling} of a convex subset of the plane by centrally symmetric convex polygons; this tiling has one polygon for each crossing point of the arrangement, which may be chosen to have unit-length edges perpendicular to the lines crossing at that point. De Bruijn~\cite{Bru-IM-81} used this construction to form aperiodic tilings (including the rhombic Penrose tiling) from overlaid families of evenly-spaced parallel lines.
More generally the dual of a \emph{weak pseudoline arrangement} (a set of curves that are the image of a line under a homeomorphism of the plane, that have at most one point of intersection per pair of curves, and that cross each other at their intersection points), may be represented as a tiling of a nonconvex simple polygon by centrally-symmetric polygons, and every such tiling arises in this way~\cite{Epp-GD-04,EppFalOvc-07}; this provides a convenient method for visualizing squaregraphs, which are the duals of triangle-free hyperbolic line arrangements~\cite{BanCheEpp-09}.

However, the tilings obtained directly from this construction may be hard to read due to having polygons with very sharp angles, as is true for instance for the tiling in Figure~\ref{fig:Ageev} (left). In unpublished work with Kevin Wortman~\cite{oarfsd}, we considered the problem of finding a combinatorially equivalent tiling that maximizes the minimum angle in the tiling (the so-called angular resolution~\cite{CarEpp-GD-06,MalPap-SJDM-94}), such as the one in Figure~\ref{fig:Ageev} (right); we showed that this optimal tiling could be found by a parametric shortest path computation in an auxiliary graph derived from the input.

Our algorithm constructs a graph in which the edges have weights that are linear functions of a parameter $\lambda$ that represents the angular resolution of the drawing. In a tiling of this type, one can define an equivalence relation on the sides of tiles in which opposite pairs of sides on the same side are equivalent; the equivalence classes form \emph{zones} of line segments that are required to be parallel and to have the same length. Our graph has one vertex per zone, and an additional start vertex that has a zero-length edge to each other vertex; the distance from the start vertex to a zone's vertex represents the adjustment in angle for the segments in the zone between an initial tiling and the optimal tiling. The constraint that the angle between two sides of a polygon be at least $\lambda$ can then be expressed by the existence of an edge between the vertices representing the zones containing the two sides, with length $\theta_i-\theta_j-\lambda$, where $\theta_i$ and $\theta_j$ are the angles formed by the two zones in the initial tiling. Similarly, the constraint that an interior angle be convex can be expressed by an edge with length $\pi+\theta_i-\theta_j$. With these vertices and edge lengths, it can be shown that there exists a tiling with angular resolution at least $\lambda$ if and only if the result of substituting that value into each edge weight function is a graph with no negative cycles. Therefore, the optimal angular resolution is the largest value of $\lambda$ giving no negative cycles. Due to the special form of the weights in the parametric graph (each weight is either constant or a constant minus $\lambda$) this parametric negative cycle detection problem can be solved in $O(n^3)$ time by an algorithm of Karp and Orlin~\cite{KarOrl-TR-80}. The translation from the tiling angular resolution problem to the parametric negative cycle detection problem and back can be performed within the same time bound.

\section{Metric embedding into stars}

There has been a large amount of interest recently within the theoretical computer science community in problems of embedding unstructured metric spaces (which may be specified, for instance, as a distance matrix) into simpler and more highly constrained metrics, with low distortion~\cite{Lin-ICM-02}. Such embeddings may be used, for instance, in approximation algorithms: one can design an approximation algorithm for the constrained class of metrics, and apply it to arbitrary metrics, incurring the distortion of the embedding as a penalty factor in the approximation ratio of the algorithm. The construction of graph spanners~\cite{Epp-HCG-00} may also be seen as an instance of this type of problem: one wishes to approximate a metric describing the shortest paths in an arbitrary graph, by a more highly constrained metric of shortest paths in a sparse graph. Most work in this area has concentrated on finding embeddings that guarantee low but non-optimal distortion; however there has also been some work on finding the best possible embedding~\cite{EppWor-WADS-09,FomLokSau-WG-09}. In particular, we describe here our work on finding optimal embeddings into star metrics, which (as in the angle optimization of the previous section) involves translating the problem into one of parametric negative cycle detection in an auxiliary graph.

\begin{figure}[t]
\centering\includegraphics[width=3in]{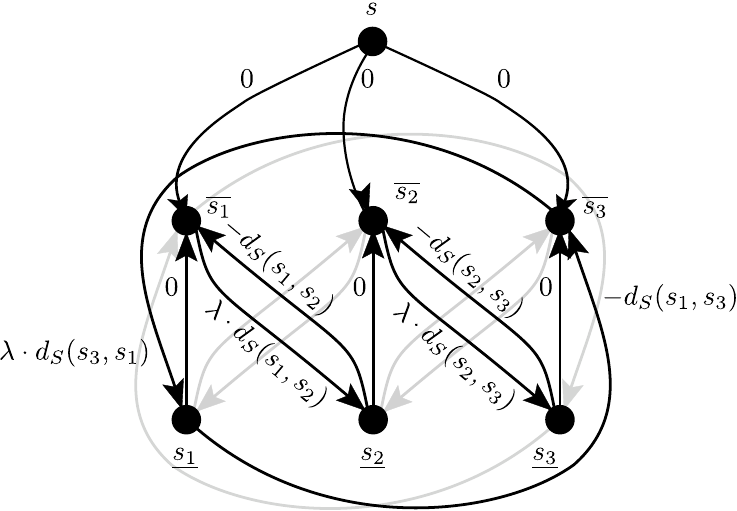}
\caption{The parametric network into which an optimal star embedding problem is translated. From~\cite{EppWor-WADS-09}.}
\label{fig:star-redux}
\end{figure}

Following~\cite{EppWor-WADS-09}, we define a \emph{star metric} to be a metric space in which there exists a distinguished point $h$ (the \emph{hub}) of the metric) such that, for every two points $p$ and $q$, $h$ lies on a shortest path from $p$ to $q$. Expressed algebraically: $d(p,q)=d(p,h)+d(h,q)$. Thus, in contrast to arbitrary $n$-point finite metric spaces which have $n(n-1)/2$ degrees of freedom, a star metric has only $n-1$ degrees of freedom. We wish to find a map from an arbitrary finite metric space (represented as a distance matrix $D$) to a star metric space (represented as a vector $H$ of distances from each point to the hub) that minimizes the distortion of the embedding; we do not require that the hub be the image of any point in the input space. By scaling $H$, if necessary, we can assume without loss of generality that the map does not decrease any distance; thus, what we seek is a vector $H$ satisfying the constraint that, for all $p$ and $q$, $D[p,q]\le H[p]+H[q]$, and minimizing the \emph{dilation}
$$\delta=\max_{p,q} \frac{H[p]+H[q]}{D[p,q]}.$$
This is a linear program: it may be rephrased as
the problem of finding $H$ and $\delta$ subject to the linear nondecreasing-distance constraints and the linear constraints that, for each $p$ and $q$, $D[p,q]\delta- H[p]-H[q]\ge 0$, and, among all tuples of $H$ and $\delta$ that satisfy the constraints, finding the tuple minimizing $\delta$. However, it may be solved more efficiently, in strongly polynomial time, by transforming it into a parametric negative cycle detection problem.

Specifically, as our paper shows, we may find the optimal star embedding using an auxiliary graph defined from the input, having two vertices $\overline{p}$ and $\underline{p}$ for each metric space point $p$, as well as a special start vertex $s$. There is an edge with length zero from $s$ to each vertex $\overline{p}$ and from each vertex $\underline{p}$ to the corresponding $\overline{p}$. In addition, for each pair of points $p$ and $q$, there is an edge of length $-D[p,q]$ from $\underline{p}$ to $\overline{q}$ and another edge  of length $\lambda D[p,q]$ from $\overline{p}$ to $\underline{q}$. As we show, the optimal dilation $\delta$ is the smallest value that can be assigned to $\lambda$ such that the resulting graph has no negative cycles, and for this value of $\lambda$ the distance $D[p]$ of $p$ from the hub in the optimal embedding may be computed as half of the difference between the two distances from $s$ to $\overline{p}$ and $\underline{p}$.

Thus, the star metric embedding problem may be solved optimally by a single parametric negative cycle detection calculation similar to that in the previous section. However, in this case the edge weight functions of the parametrized network no longer have the special form needed to apply the Karp--Orlin algorithm; instead, we developed a more general algorithm that solves any parametric negative cycle detection problem in strongly polynomial time. The basic idea of the algorithm is to use a matrix squaring method of Savage~\cite{Sav-PhD-77} to calculate the parametrized function representing the weights of paths between each pair of vertices; after $i$ iterations of squaring a matrix whose entries represent these functions, they will correctly describe the lengths of the shortest path between each pair of vertices among paths constrained to have at most $2^i$ hops. After each matrix-squaring step, the entries of the matrix will not be linear functions but rather more general piecewise linear concave functions; we perform a binary search among the set of breakpoints of these functions to narrow down the range of values of $\lambda$ within which the optimal value must lie, allowing us to simplify the functions to be linear once again. Each binary search step involves applying the Bellman-Ford algorithm to determine whether a specific value of $\lambda$ causes the parametrized graph to have a negative cycle. The negative cycle detection algorithm resulting from this parametric search procedure has total running time $O(n^3\log^2 n)$, which is therefore also the time for using this approach to solve the optimal star metric embedding problem.

\section{Conclusions}

As we have seen, a graph-theoretic point of view can be useful in many algorithmic problems that do not, initially, seem to have much to do with graphs. The graph problems occurring in the solutions of the geometry problems discussed here are mostly well-known and classical: they include the maximum independent set and maximum clique problems, maximum cardinality matching, shortest paths, and minimum cost flow. However, sometimes these problems occur in a somewhat different form than they have been studied elsewhere in the literature: the star metric embedding problem, for instance, required parametric negative cycle detection at a level of generality that had previously not been considered.

Special classes of graphs, and specialized algorithms that take advantage of these special classes, seem to be ubiquitous in this area. In the examples here, we have seen natural applications of trees, maximal outerplanar graphs, squaregraphs and kinggraphs, bipartite graphs, unit disk graphs, apex graphs, and 3-regular bridgeless graphs. The simple coloring algorithm for maximal outerplanar graphs described in the section on the art gallery theorem generalizes to chordal graphs (maximal outerplanar graphs are chordal) and to perfectly orderable graphs~\cite{Chv-TiPG-84}.

There are undoubtedly many more applications of graph-theoretic concepts in computational geometry waiting to be discovered, and it is important that the lines of communication remain open between the graph algorithm and computational geometry communities, so that computational geometers will know where to find the algorithms they need and so that graph algorithms researchers can focus their efforts on the problems and graph classes with the greatest benefit in geometric applications.

\subsubsection*{Acknowledgements}

This work was supported in part by NSF grant
0830403 and by the Office of Naval Research under grant
N00014-08-1-1015.

\raggedright
\bibliographystyle{abbrv}
\bibliography{geomgraphs}

\begin{thebibliography}{10}

\bibitem{AgaEfrSha-SJC-00}
P.~K. Agarwal, A.~Efrat, and M.~Sharir.
\newblock {Vertical decomposition of shallow levels in 3-dimensional
  arrangements and its applications}.
\newblock {\em SIAM J. Comput.}, 29(3):912{--}953, 2000.

\bibitem{Age-DM-96}
A.~A. Ageev.
\newblock {A triangle-free circle graph with chromatic number 5}.
\newblock {\em Discrete Mathematics}, 152:295{--}298, 1996.

\bibitem{AggImaKat-Algs-91}
A.~Aggarwal, H.~Imai, N.~Katoh, and S.~Suri.
\newblock {Finding $k$ points with minimum diameter and related problems}.
\newblock {\em J. Algorithms}, 12(1):38{--}56, 1991.

\bibitem{AjtChvNew-TaPoC-82}
M.~Ajtai, V.~Chv{\'a}tal, M.~Newborn, and E.~Szemer{\'e}di.
\newblock {Crossing-free subgraphs}.
\newblock In {\em Theory and Practice of Combinatorics}, pages 9{--}12.
  North-Holland Mathematics Studies, vol. 60, 1982.

\bibitem{ArkHelMit-VC-96}
E.~M. Arkin, M.~Held, J.~S.~B. Mitchell, and S.~S. Skiena.
\newblock {Hamiltonian triangulations for fast rendering}.
\newblock {\em The Visual Computer}, 12(9):429{--}444, 1996.

\bibitem{BanCheEpp-09}
H.-J. Bandelt, V.~Chepoi, and D.~Eppstein.
\newblock {Combinatorics and geometry of finite and infinite squaregraphs}.
\newblock Electronic preprint arxiv:0905.4537, 2009.

\bibitem{BieBosDem-Algs-01}
T.~C. Biedl, P.~Bose, E.~D. Demaine, and A.~Lubiw.
\newblock {Efficient algorithms for Petersen's matching theorem}.
\newblock {\em J. Algorithms}, 38:110{--}134, 2001.

\bibitem{BreKir-CGTA-98}
H.~Breu and D.~G. Kirkpatrick.
\newblock {Unit disk graph recognition is NP-hard}.
\newblock {\em Computational Geometry Theory and Applications},
  9(1{--}2):3{--}24, 1998.

\bibitem{CarEpp-GD-06}
J.~Carlson and D.~Eppstein.
\newblock {Trees with convex faces and optimal angles}.
\newblock In M.~Kaufmann and D.~Wagner, editors, {\em Proc. 14th Int. Symp.
  Graph Drawing}, volume 4372 of {\em Lecture Notes in Computer Science}, pages
  77{--}88. Springer-Verlag, 2006.

\bibitem{Cha-DCG-91}
B.~Chazelle.
\newblock {Triangulating a simple polygon in linear time}.
\newblock {\em Discrete {\&} Computational Geometry}, 6(1):485{--}524, 1991.

\bibitem{CheIyeKas-TSE-88}
Y.~Cheng, S.~S. Iyengar, and R.~L. Kashyap.
\newblock {A new method of image compression using irreducible covers of
  maximal rectangles}.
\newblock {\em IEEE Trans. Software Engineering}, 14(5):651{--}658, 1988.

\bibitem{CheDraVax-SODA-02}
V.~Chepoi, F.~Dragan, and Y.~Vax{\`e}s.
\newblock {Center and diameter problem in planar quadrangulations and
  triangulations}.
\newblock In {\em Proc. 13th Annu. ACM{--}SIAM Symp. on Discrete Algorithms
  (SODA 2002)}, pages 346{--}355, 2002.

\bibitem{Chv-JCTB-75}
V.~Chv{\'a}tal.
\newblock {A combinatorial theorem in plane geometry}.
\newblock {\em Journal of Combinatorial Theory, Series B}, 18:39{--}41, 1975.

\bibitem{Chv-TiPG-84}
V.~Chv{\'a}tal.
\newblock {Perfectly orderable graphs}.
\newblock In C.~Berge and V.~Chv{\'a}tal, editors, {\em Topics in Perfect
  Graphs}, volume~21 of {\em Annals of Discrete Mathematics}, pages 63{--}68.
  North-Holland, Amsterdam, 1984.

\bibitem{ClaColJoh-DM-90}
B.~N. Clark, C.~J. Colbourn, and D.~S. Johnson.
\newblock {Unit disk graphs}.
\newblock {\em Discrete Mathematics}, 86:165{--}177, 1990.

\bibitem{Bru-IM-81}
N.~G. de~Bruijn.
\newblock {Algebraic theory of Penrose's non-periodic tilings of the plane}.
\newblock {\em Indagationes Mathematicae}, 43:38{--}66, 1981.

\bibitem{Dee-SIGGRAPH-95}
M.~Deering.
\newblock {Geometry compression}.
\newblock In {\em Proc. 22nd Conf. Computer Graphics and Interactive Techniques
  (SIGGRAPH)}, pages 13{--}20, 1995.

\bibitem{Dey-DCG-98}
T.~K. Dey.
\newblock {Improved bounds for planar $k$-sets and related problems}.
\newblock {\em Discrete {\&} Computational Geometry}, 19(3):373{--}382, April
  1998.

\bibitem{BatEadTam-98}
G.~Di~Battista, P.~Eades, R.~Tamassia, and I.~G. Tollis.
\newblock {\em {Graph Drawing: Algorithms for the Visualization of Graphs}}.
\newblock Prentice Hall, 1998.

\bibitem{Dia-PhD-09}
P.~D{\'\i}az-Guti{\'e}rrez.
\newblock {\em {Using graph algorithms for geometry processing on surfaces}}.
\newblock PhD thesis, Univ. of California, Irvine, 2009.

\bibitem{Eng-DAM-09}
K.~Engel.
\newblock {Optimal matrix-segmentation by rectangles}.
\newblock {\em Discrete Applied Mathematics}, 157(9):2015{--}2030, 2009.

\bibitem{Epp-DCG-98}
D.~Eppstein.
\newblock {Geometric lower bounds for parametric matroid optimization}.
\newblock {\em Discrete {\&} Computational Geometry}, 20:463{--}476, 1998.

\bibitem{Epp-HCG-00}
D.~Eppstein.
\newblock {Spanning trees and spanners}.
\newblock In J.-R. Sack and J.~Urrutia, editors, {\em Handbook of Computational
  Geometry}, chapter~9, pages 425{--}461. Elsevier, 2000.

\bibitem{Epp-GD-04}
D.~Eppstein.
\newblock {Algorithms for drawing media}.
\newblock In {\em Proc. 12th Int. Symp. Graph Drawing}, volume 3383 of {\em
  Lecture Notes in Computer Science}, pages 173{--}183. Springer-Verlag, 2005.

\bibitem{Epp-JGAA-07}
D.~Eppstein.
\newblock {The traveling salesman problem for cubic graphs}.
\newblock {\em J. Graph Algorithms and Applications}, 11(1):61{--}81, 2007.

\bibitem{Epp-TALG-bgig}
D.~Eppstein.
\newblock {Testing bipartiteness of geometric intersection graphs}.
\newblock {\em ACM Trans. Algorithms}, 5(2):15, 2009.

\bibitem{EppEri-DCG-94}
D.~Eppstein and J.~Erickson.
\newblock {Iterated nearest neighbors and finding minimal polytopes}.
\newblock {\em Discrete {\&} Computational Geometry}, 11(3):321{--}350, April
  1994.

\bibitem{EppFalOvc-07}
D.~Eppstein, J.-C. Falmagne, and S.~Ovchinnikov.
\newblock {\em {Media Theory}}.
\newblock Springer-Verlag, 2007.

\bibitem{EppGop-EG-04}
D.~Eppstein and M.~Gopi.
\newblock {Single-strip triangulation of manifolds with arbitrary topology}.
\newblock {\em Eurographics Forum}, 23(3):371{--}379, 2004.
\newblock Proc. 25th Conf. Eur. Assoc. for Computer Graphics (EuroGraphics
  2004).

\bibitem{EppMum-WADS-09}
D.~Eppstein and E.~Mumford.
\newblock {Orientation-constrained rectangular layouts}.
\newblock In {\em Proc. Algorithms and Data Structures Symposium (WADS 2009)},
  volume 5664 of {\em Lecture Notes in Computer Science}, pages 266{--}277.
  Springer-Verlag, 2009.

\bibitem{EppMumSpe-SCG-09}
D.~Eppstein, E.~Mumford, B.~Speckmann, and K.~A.~B. Verbeek.
\newblock {Area-universal rectangular layouts}.
\newblock In {\em Proc. 25th ACM Symp. Computational Geometry}, pages
  267{--}276, 2009.

\bibitem{oarfsd}
D.~Eppstein and K.~Wortman.
\newblock {Optimal angular resolution for face-symmetric drawings}.
\newblock Electronic preprint arxiv:0907.5474, 2009.

\bibitem{EppWor-WADS-09}
D.~Eppstein and K.~Wortman.
\newblock {Optimal embedding into star metrics}.
\newblock In {\em Proc. Algorithms and Data Structures Symposium (WADS 2009)},
  volume 5664 of {\em Lecture Notes in Computer Science}, pages 290{--}301.
  Springer-Verlag, 2009.

\bibitem{EvaSkiVar-V-96}
F.~Evans, S.~S. Skiena, and A.~Varshney.
\newblock {Optimizing triangle strips for fast rendering}.
\newblock In {\em Proc. 7th IEEE Conf. Visualization}, pages 319{--}326, 1996.

\bibitem{FerSanSkl-CVGIP-84}
L.~Ferrari, P.~V. Sankar, and J.~Sklansky.
\newblock {Minimal rectangular partitions of digitized blobs}.
\newblock {\em Computer Vision, Graphics, and Image Processing},
  28(1):58{--}71, 1984.

\bibitem{Fis-JCTB-78}
S.~Fisk.
\newblock {A short proof of Chv{\'a}tal's watchman theorem}.
\newblock {\em Journal of Combinatorial Theory, Series B}, 24(3):374, 1978.

\bibitem{FomLokSau-WG-09}
F.~Fomin, D.~Lokshtanov, and S.~Saurabh.
\newblock {An exact algorithm for minimum distortion embedding}.
\newblock In {\em Proc. 35th Int. Worksh. Graph-Theoretic Concepts in Computer
  Science}, Lecture Notes in Computer Science. Springer-Verlag, 2009.
\newblock This proceedings.

\bibitem{GarTam-GD-96}
A.~Garg and R.~Tamassia.
\newblock {A new minimum cost flow algorithm with applications to graph
  drawing}.
\newblock In {\em Proc. 6th Int. Symp. Graph Drawing}, volume 1190 of {\em
  Lecture Notes in Computer Science}, pages 201{--}216. Springer-Verlag, 1997.

\bibitem{HanHubLip-CT-02}
S.~Hannenhalli, E.~Hubbell, R.~Lipshutz, and P.~A. Pevzner.
\newblock {Combinatorial algorithms for design of DNA arrays}.
\newblock In {\em Chip Technology}, volume~77 of {\em Advances in Biochemical
  Engineering/Biotechnology}, pages 1{--}19. Springer-Verlag, 2002.

\bibitem{HerSur-Algs-91}
J.~Hershberger and S.~Suri.
\newblock {Finding tailored partitions}.
\newblock {\em J. Algorithms}, 12(3):431{--}463, 1991.

\bibitem{HopKar-SJC-73}
J.~E. Hopcroft and R.~M. Karp.
\newblock {An $n^{5/2}$ algorithm for maximum matchings in bipartite graphs}.
\newblock {\em SIAM J. Comput.}, 2(4):225{--}231, 1973.

\bibitem{ImaAsa-SJC-86}
H.~Imai and T.~Asano.
\newblock {Efficient algorithms for geometric graph search problems}.
\newblock {\em SIAM J. Comput.}, 15:478{--}494, 1986.

\bibitem{ImaIwa-SIGAL-90}
H.~Imai and K.~Iwano.
\newblock {Efficient sequential and parallel algorithms for planar minimum cost
  flow}.
\newblock In {\em Proc. Int. Symp. Algorithms}, volume 450 of {\em Lecture
  Notes in Computer Science}, pages 21{--}30. Springer-Verlag, 1990.

\bibitem{IwaNak-COCOON-07}
K.~Iwama and T.~Nakashima.
\newblock {An improved exact algorithm for cubic graph TSP}.
\newblock In {\em Proc. 13th Int. Conf. Computing and Combinatorics (COCOON)},
  volume 4598 of {\em Lecture Notes in Computer Science}, pages 108{--}117.
  Springer-Verlag, 2007.

\bibitem{JunMut-04}
M.~Junger and P.~Mutzel.
\newblock {\em {Graph Drawing Software}}.
\newblock Springer-Verlag, 2004.

\bibitem{KahKlaKle-SJDM-83}
J.~Kahn, M.~Klawe, and D.~Kleitman.
\newblock {Traditional galleries require fewer watchmen}.
\newblock {\em SIAM Journal on Algebraic and Discrete Methods},
  4(2):194{--}206, 1983.

\bibitem{Kal-EJC-09}
T.~Kalinowski.
\newblock {A dual of the rectangle-segmentation problem for binary matrices}.
\newblock {\em Electronic J. Combinatorics}, 16(1):R89, 2009.

\bibitem{KarOrl-TR-80}
R.~M. Karp and J.~B. Orlin.
\newblock {Parametric Shortest Path Algorithms with an Application to Cyclic
  Staffing}.
\newblock Technical Report OR 103-80, MIT Operations Research Center, 1980.

\bibitem{Kon-MeFL-31}
D.~K{\H{o}}nig.
\newblock {Gr{\'a}fok {\'e}s m{\'a}trixok}.
\newblock {\em Matematikai {\'e}s Fizikai Lapok}, 38:116{--}119, 1931.

\bibitem{Lei-83}
T.~Leighton.
\newblock {\em {Complexity Issues in VLSI}}.
\newblock Foundations of Computing Series. MIT Press, Cambridge, MA, 1983.

\bibitem{LiZha-IROS-05}
G.~Li and H.~Zhang.
\newblock {A rectangular partition algorithm for planar self-assembly}.
\newblock In {\em Proc. IEEE/RSJ Int. Conf. Intelligent Robots and Systems},
  pages 3213{--}3218, 2005.

\bibitem{Lin-ICM-02}
N.~Linial.
\newblock {Finite metric spaces{--}combinatorics, geometry and algorithms}.
\newblock In {\em Proc. International Congress of Mathematicians, Beijing},
  volume~3, pages 573{--}586, 2002.

\bibitem{Lip-Nw-83}
W.~Lipski, Jr.
\newblock {Finding a Manhattan path and related problems}.
\newblock {\em Networks}, 13:399{--}409, 1983.

\bibitem{Lip-IPL-84}
W.~Lipski, Jr.
\newblock {An $O(n\log n)$ Manhattan path algorithm}.
\newblock {\em Information Processing Letters}, 19:99{--}102, 1984.

\bibitem{LipLodLuc-FI-79}
W.~Lipski, Jr., E.~Lodi, F.~Luccio, C.~Mugnai, and L.~Pagli.
\newblock {On two-dimensional data organization II}.
\newblock {\em Fundamenta Informaticae}, 2:245{--}260, 1979.

\bibitem{MalPap-SJDM-94}
S.~Malitz and A.~Papakostas.
\newblock {On the angular resolution of planar graphs}.
\newblock In {\em SIAM J. Discrete Mathematics}, volume~7, pages 172{--}183,
  1994.

\bibitem{Mil-SJC-95}
G.~Miller.
\newblock {Flow in planar graphs with multiple sources and sinks}.
\newblock {\em SIAM J. Comput.}, 24(5):1002{--}1017, 1995.

\bibitem{Mum-PhD-08}
E.~Mumford.
\newblock {\em {Drawing Graphs for Cartographic Applications}}.
\newblock PhD thesis, Technische Universiteit Eindhoven, 2008.

\bibitem{NisRah-04}
T.~Nishizeki and M.~S. Rahman.
\newblock {\em {Planar Graph Drawing}}.
\newblock World Scientific, 2004.

\bibitem{Oht-ISCAS-82}
T.~Ohtsuki.
\newblock {Minimum dissection of rectilinear regions}.
\newblock In {\em Proc. IEEE Int. Symp. Circuits and Systems}, pages
  1210{--}1213, 1982.

\bibitem{ORo-87}
J.~O'Rourke.
\newblock {\em {Art Gallery Theorems and Algorithms}}.
\newblock Oxford University Press, 1987.

\bibitem{Pat-CAD-77}
K.~Patel.
\newblock {Computer-aided decomposition of geometric contours into standardized
  areas}.
\newblock {\em Computer-Aided Design}, 9(3):199{--}203, 1977.

\bibitem{Pet-AM-91}
J.~P.~C. Petersen.
\newblock {Die theorie der regularen graphs}.
\newblock {\em Acta Mathematica}, 15:193{--}220, 1891.

\bibitem{Rai-GR-34}
E.~Raisz.
\newblock {The rectangular statistical cartogram}.
\newblock {\em Geographical Review}, 24(2):292{--}296, 1934.

\bibitem{Kei-HCG-99}
J.-R. Sack and J.~Urrutia.
\newblock {Polygon decomposition}.
\newblock In J.-R. Sack and J.~Urrutia, editors, {\em Handbook of Computational
  Geometry}, pages 491{--}518. Elsevier, 1999.

\bibitem{Sav-PhD-77}
C.~Savage.
\newblock {\em {Parallel Algorithms for Graph Theoretic Problems}}.
\newblock PhD thesis, University of Illinois, Urbana-Champaign, 1977.

\bibitem{ShaHoe-FOCS-75}
M.~I. Shamos and D.~Hoey.
\newblock {Closest-point problems}.
\newblock In {\em Proc. 16th IEEE Symp. Foundations of Computer Science}, pages
  151{--}162, 1975.

\bibitem{Tai-PM-84}
P.~G. Tait.
\newblock {Listing's {\it Topologie}}.
\newblock {\em Philosophical Magazine (5th ser.)}, 17:30{--}46, 1884.

\bibitem{Tam-SJC-87}
R.~Tamassia.
\newblock {On embedding a graph in the grid with the minimum number of bends}.
\newblock {\em SIAM J. Comput.}, 16(3):421{--}444, 1987.

\bibitem{Tut-JLMS-46}
W.~T. Tutte.
\newblock {On Hamiltonian circuits}.
\newblock {\em Journal of the London Mathematical Society (2nd ser.)},
  21(2):98{--}101, 1946.
\newblock Reprinted in {\it Scientific Papers}, Vol. II, pp. 85{--}98.

\bibitem{Vai-SJC-89}
P.~M. Vaidya.
\newblock {Geometry helps in matching}.
\newblock {\em SIAM J. Comput.}, 18(6):1201{--}1225, 1989.

\bibitem{KreSpe-CGTA-07}
M.~van Kreveld and B.~Speckmann.
\newblock {On rectangular cartograms}.
\newblock {\em Computational Geometry Theory and Applications},
  37(3):175{--}187, 2007.

\bibitem{XiaHelMit-3DG-99}
X.~Xiang, M.~Held, and J.~S.~B. Mitchell.
\newblock {Fast and effective stripification of polygonal surface models}.
\newblock In {\em Proc. Symp. Interactive 3D Graphics}, pages 71{--}78, 1999.

\end{thebibliography}
\end{document}